\documentclass [11pt]{article}%
\usepackage{epsf} 
\usepackage{amsmath}
\usepackage{amssymb}
\usepackage{epsfig}
\usepackage{latexsym}
\usepackage{amsfonts}
\usepackage{graphicx}%
\usepackage{varioref}
\usepackage{ifthen}
\begin{document}
\parindent 0mm 
\setlength{\parskip}{\baselineskip} 
\thispagestyle{empty}
\pagenumbering{arabic} 
\setcounter{page}{0}
\mbox{ }
\rightline{UCT-TP-277/09}
\newline\rightline{August 2009}
\newline\rightline{revised December 2009}
\newline
\vspace{0.2cm}
\begin{center}
{\Large {\bf Charmonium in the vector channel at finite temperature from QCD sum rules\textbf{}}}
{\LARGE \footnote{{\LARGE {\footnotesize Supported in part by FONDECYT 1060653, 1095217 and 7080120 (Chile), Centro de Estudios Subatomicos (Chile), and NRF (South Africa).}}}}
\end{center}
\vspace{.1cm}
\begin{center}
{\bf C. A. Dominguez}$^{(a)-(b)}$,  {\bf M. Loewe},$^{(c)}$, {\bf J. C. Rojas}$^{(d)}$,\\
 {\bf Y. Zhang}$^{(a)}$
\end{center}
\begin{center}
$^{(a)}$Centre for Theoretical Physics and Astrophysics\\[0pt]University of
Cape Town, Rondebosch 7700, South Africa

$^{(b)}$ Department of Physics, Stellenbosch University, Stellenbosch 7600, South Africa

$^{(c)}$Facultad de F\'{i}sica, Pontificia Universidad Cat\'{o}lica de Chile, Casilla 306, Santiago 22, Chile

$^{(d)}$ Departamento de F\'{i}sica,  Universidad Cat\'{o}lica del Norte, Casilla 1280, Antofagasta, Chile\end{center}\vspace{0.3cm}
\begin{center}
\textbf{Abstract}
\end{center}
Thermal Hilbert moment QCD sum rules are used to obtain the temperature dependence of the hadronic parameters of charmonium in the vector channel, i.e. the $J$/$\psi$ resonance mass, coupling (leptonic decay constant), total width, and continuum threshold. The continuum threshold $s_0$, which signals the end of the resonance region and the onset of perturbative QCD (PQCD), behaves as in all other hadronic channels, i.e. it decreases with increasing temperature until it reaches 
the PQCD threshold $s_0 = 4\, m_Q^2$, with $m_Q$ the charm quark mass,
 at $T\simeq \, 1.22 \, T_c$. The rest of the hadronic parameters behave  very differently from those of light-light and heavy-light quark systems. The $J$/$\psi$ mass is essentially constant in a wide range of temperatures, while the total width grows with temperature up to $T \simeq \, 1.04 \,T_c$ beyond which it decreases sharply with increasing T. The resonance coupling is also initially constant and then begins to increase monotonically around $T \simeq T_c$. This  behaviour  of the total width and of the leptonic decay constant provides a strong indication that the $J$/$\psi$ resonance might survive beyond the critical temperature for deconfinement.\\

\newpage
\bigskip
\noindent
\section{Introduction}
\noindent
A successful quantum field theory framework to extract hadronic information from QCD analytically is that of QCD sum rules \cite{QCDSR}. This technique  is based on the Operator Product Expansion (OPE) of current correlators beyond perturbation theory, and on Cauchy's theorem in the complex energy  plane (quark-hadron duality). This program was first extended to finite temperature  in \cite{BOCH}. It is based on two basic assumptions, (a) that the OPE continues to be valid, with the vacuum condensates developing a temperature dependence, and (b) that no thermal singularities appear in the complex energy  plane, other than on the real axis, i.e. the notion of quark-hadron duality also remains valid. Field theory evidence in support of these assumptions was provided later in \cite{CAD1}. Numerous applications of QCD sum rules at finite temperature have been made over the years \cite{VARIOUS}-\cite{HL}, leading to the following scenario for light-light and heavy-light quark hadrons. (i) As the temperature increases, hadronically stable particles develop a non-zero  width, and resonances become broader, diverging at a critical temperature interpreted as the  deconfinement temperature ($T_c$). This width is a result of particle absorption in the thermal bath, and resonance broadening at finite temperature was first proposed in \cite{CAD0}. (ii) Above the resonance region the  continuum threshold  in hadronic spectral functions, i.e. the onset of  perturbative QCD (PQCD), decreases monotonically with increasing temperature. In other words, as $T \rightarrow T_c$ hadrons melt  disappearing from the spectrum, which then becomes  smooth. (iii) Additional support for this picture is provided by the behaviour of hadronic couplings, or leptonic decay constants, which approach zero as $T \rightarrow T_c$. Also, hadronic and electromagnetic mean-squared radii diverge at $T_c$ \cite{radii} indicating deconfinement. On a separate issue, QCD sum rules in the axial-vector channel have provided (analytical) evidence for the (almost) equality of the critical temperatures for deconfinement and chiral-symmetry restoration \cite{CAD3}. Contrary to this revealing behaviour of widths and couplings, the mass does not appear to  offer any relevant  information about deconfinement.
In fact, in most cases it increases or decreases, sometimes   slightly, with increasing $T$.
Conceptually, given either the emergence or the broadening of an existing width, together with its divergence at $T_c$, the concept of mass looses most of its meaning.\\
At this stage it must be pointed out that in the framework of QCD sum rules the critical temperature for deconfinement, referred to above, is only a phenomenological parameter. It is the temperature at which the resonance couplings and the continuum threshold approach zero, and the widths increase sharply, for light-light and heavy-light quark correlators. Hence, it need not coincide numerically with e.g. the critical temperature obtained in lattice QCD \cite{lattice}, which is defined differently. In fact, results from QCD sum rules lead to  values of $T_c$ somewhat lower than those from lattice QCD. Hence, comparisons between different frameworks should be made in terms of the dependence of parameters on the ratio $T/T_c$.\\
Turning to heavy-heavy quark hadrons at finite $T$, in principle one would expect them to behave  differently from the light-light and heavy-light quark systems for the following reasons. The $T$ dependence of the latter in the PQCD sector is dominated by the spectral function in the time-like region, the so-called annihilation term, which is anyway relatively unimportant in relation to the light quark condensate contribution.  The PQCD spectral function in  the space-like region (scattering term) is highly suppressed. For heavy-heavy quark systems this is not the case; the scattering term becomes increasingly important with increasing temperature while the annihilation term only contributes near threshold. In the non-perturbative QCD sector of light-light and heavy-light quark correlators, the driving term in the OPE is the light quark condensate. This term is responsible for the behaviour of the continuum threshold, as it follows that $s_0(T)/s_0(0) \simeq \langle\langle \bar{q} q\rangle\rangle/\langle \bar{q} q\rangle$ \cite{CAD2}-\cite{radii}. This scaling relation is rather important as the thermal light quark condensate is the order parameter for chiral symmetry restoration. In contrast, for heavy-heavy quark correlators the leading power correction in the OPE is that of the gluon condensate, which has a very different temperature behaviour. This expectation about a different temperature behaviour of heavy-heavy quark systems is shown here to hold for the case of the $J/\psi$. Using Hilbert moment QCD sum rules we obtain the $J/\psi$ hadronic parameters at finite temperature.  With the exception of the continuum threshold, we find a very different behaviour from that of light-light and heavy-light quark systems.  The  continuum threshold, $s_0(T)$, does decrease with increasing $T$, being driven by the gluon condensate and the PQCD spectral function in the space-like region, until it reaches the PQCD threshold $s_0 = 4\, m_Q^2$, where $m_Q$ is the charm quark mass, at $T\simeq \, 1.22\, T_c$. The $J/\psi$ mass remains basically constant. The width and the coupling are almost independent of $T$ up to $T \simeq \, 0.8\, T_c$ where the width begins to increase substantially, but then above $T \simeq \, 1.04 \, T_c$ it starts to decrease sharply, and the coupling  increases also sharply. This behaviour, which can mostly be traced to that of the PQCD spectral function in the space-like region, points to the survival of the $J/\psi$ resonance above the deconfinement temperature. However, the QCD sum rules have no longer solutions for the hadronic parameters once the continuum threshold reaches the value $s_0|_{min} = M^2_{J/\psi}$, as there is no longer any support for the integrals. Hence, the temperature range explored with this technique does not extend  beyond  $T \simeq \, 1.22 \, T_c$. Given the importance of the PQCD spectral function in the space-like region, it should be realized that non-relativistic approaches to charmonium at finite temperature will most probably miss this contribution. In fact, the complex energy plane in the non-relativistic case would only have one cut along the positive real axis, which would correspond to the time-like (annihilation) region of PQCD. The space-like contribution  ($q^2 = (\omega^2 - |\bf{q}|^2) \leq 0$) in the form of a cut in the energy plane centered at the origin for $- |\bf{q}| \leq \omega \leq |\bf{q}|$, would not be present in the non-relativistic case.
\section{Hilbert moment QCD sum rules}
\noindent
We consider the correlator of the heavy-heavy quark vector current at finite temperature
\begin{eqnarray}
\Pi_{\mu\nu} (q^{2},T)   &=& i \, \int\; d^{4} \, x \; e^{i q x} \; \;\theta(x_0)\;
<<|[ V_{\mu}(x) \;, \; V_{\nu}^{\dagger}(0)]|>> \nonumber \\ [.3cm]
&=& -(g_{\mu\nu} q^2 - q_\mu q_\nu) \Pi(q^2,T)  \; ,
\end{eqnarray}
where $V_\mu(x) = : \bar{Q}(x) \gamma_\mu Q(x):$, and $Q(x)$ is the heavy (charm) quark field.
The vacuum to vacuum matrix element above is the Gibbs average
\begin{equation}
<< A \cdot B>> = \sum_n exp(-E_n/T) <n| A \cdot B|n> / Tr (exp(-H/T)) \;,
\end{equation}
where $|n>$ is any complete set of eigenstates of the (QCD) Hamiltonian. We shall adopt the quark-gluon basis, as this allows for the standard QCD sum rule program to be smoothly extended to finite temperature \cite{CAD1}.\\ 
The imaginary part of the vector correlator  in perturbative QCD (PQCD) at finite temperature, $Im \; \Pi(q^2,T)$, involves two pieces, one in the time-like region ($q^2 \geq 4 m_Q^2$), $Im \; \Pi_a(q^2,T)$, which survives at T=0, and one in the space-like region ($q^2 \leq 0$), $Im \; \Pi_s(q^2,T)$, which vanishes at T=0. A straightforward calculation in the time-like region, to leading order in PQCD, gives 

\begin{equation}
\frac{1}{\pi}\, Im \,\Pi_a(q^2,T) =  \frac{3}{16 \pi^2}\;\int_{-v}^{v}\;dx \;(1-x^2) \left[1 - n_F\left(\frac{|\mathbf{q}| x + \omega}{2 T}\right)
- n_F\left( \frac{|\mathbf{q}| x -\omega}{2 T} \right)\right] \;,
\end{equation}

where $ v^2 = 1 - 4 m_Q^2/q^2$, $m_Q$ is the heavy quark mass, $q^2 = \omega^2 - \mathbf{q}^2 \geq 4 m_Q^2$, and  $n_F(z) = (1+e^z)^{-1}$ is the Fermi thermal function.
In the rest frame of the thermal bath, $|\mathbf{q}| \rightarrow 0$, the above result reduces to
\begin{equation}
\frac{1}{\pi}\, Im \,\Pi_a(\omega,T) =  \frac{1}{8 \pi^2}\; v (3 - v^2) \left[1 - 2 n_F (\omega/2 T)\right] \; \theta(\omega - 2 m_Q)
\;.
\end{equation}
The quark mass is assumed independent of $T$, which is a good approximation for temperatures below 200 MeV \cite{mQ}. As is customary in all QCD sum rule analyses at finite temperature, only the leading order in the strong coupling will be considered here.  One reason is that the temperature introduces an additional scale, and this problem is not yet  fully understood. More importantly, though,
results in this framework are not intended to be of high precission, as the T-dependence of hadronic parameters will probably never be  measured with great accuracy (some, like e.g. the leptonic decay couplings, may not even be measured at all at finite $T$). For this reason one normally determines in this framework the ratio of hadronic parameters at finite and at zero $T$ as a function of $T/T_c$.\\

The calculation of the PQCD piece in the space-like region, the so-called scattering term, is more involved as the limit $|\mathbf{q}| \rightarrow 0$ must be taken with extreme care. In fact, in the complex energy plane, and in the space-like region  the correlator $\Pi(q^2)$,  Eq.(1), has a cut centered at the origin and extending between $\omega = -|\mathbf{q}|$ and $\omega = | \mathbf{q}|$. In the rest frame of the thermal bath this cut shrinks to zero and produces a delta function $\delta(\omega^2)$ in the imaginary part of $\Pi(q^2)$. Details of this calculation are left for the Appendix; the result is
\begin{equation}
\frac{1}{\pi}\, Im \,\Pi_s(\omega,T) =  \frac{2}{ \pi^2}\; m_Q^2\; \delta(\omega^2)\left[   n_F \left(\frac{m_Q}{T}\right)\;+ \frac{2\, T^2}{m_Q^2} \;\int_{ m_Q/T }^{\infty} y \,n_F (y)\, dy \;\right]
\;.
\end{equation}
To complete the evaluation of the vector correlator in QCD, we will add later the leading power correction in the OPE, which in this case is given in terms of the gluon condensate $\langle \langle0| \alpha_s G^2 |0\rangle\rangle$.\\
Turning to the hadronic representation of the vector correlator we shall, as usual, parametrize it in terms of the ground state resonance, i.e. the $J$/$\psi$, followed   by a continuum given by PQCD after a threshold $s_0 > M^2_{V}$. This ansatz is even a better approximation at finite temperature, as $s_0(T)$ decreases monotonically with increasing $T$. Considering first the zero width approximation, the hadronic spectral function is given by
\begin{eqnarray}
\frac{1}{\pi}\, Im \,\Pi(s,T)|_{HAD} &=&  \frac{1}{\pi} Im \,\Pi(s,T)|_{RES}\; \theta(s_0 - s) +  \frac{1}{\pi} Im \,\Pi(s,T)|_{PQCD} \;\theta(s - s_0)
\nonumber \\ [.3cm]
&=& 2 \, f_V^2(T)  \, \delta(s - M_V^2(T)) \,+  \frac{1}{\pi} Im \,\Pi(s,T)_{a} \;\theta(s - s_0)\;,
\end{eqnarray}
where $s \equiv q^2 = \omega^2 - \mathbf{q}^2$, and the leptonic decay constant is defined as 
\begin{equation}
<0| V_\mu(0) | V(k)> = \sqrt{2}\; M_V \;f_V \;\epsilon_\mu \; .
\end{equation}
Next, considering a finite (total) width  the following replacement will be understood
\begin{equation}
\delta(s- M_V^2(T)) \Longrightarrow const \; \frac{1}{(s-M_V^2(T))^2 + M_V^2(T) \Gamma_V^2(T)}\; ,
\end{equation}
where the constant is fixed by requiring equality of areas, e.g. if the integration is in the interval $(0 -\infty)$ then $ const =  M_V(T) \Gamma_V(T)/\pi$. To complete the hadronic parametrization one needs to consider the hadronic scattering term due to the current scattering off  heavy-light quark pseudoscalar mesons (D-mesons). This contribution is given by \cite{BOCH}

\begin{equation}
\frac{1}{\pi}\, Im \,\Pi_s(\omega,T)|_{HAD} =  \frac{2}{ 3 \pi^2}\; M_D^2\;\delta(\omega^2) \left[ n_B\left(\frac{M_D}{T}\right) + \frac{2 \,T^2}{M_D^2}\;\int_{m_D/T }^{\infty} y \,n_B(y)  \, dy \right] 
\;,
\end{equation}
where $n_B(z) = (1-e^z)^{-1}$ is the Bose thermal function.
As shown in \cite{HL} heavy-light pseudoscalar mesons deconfine at the critical temperature, which in that application was $T_c \simeq 100 - 110\; \mbox{MeV}$. Since, a posteriori, the critical temperature for the $J/\psi$ is much higher, this scattering term should not contribute.  More importantly, though, it is easy to see by comparing Eq.(9) with Eq.(5) that this hadronic scattering term is exponentially suppressed; it is, in fact, two to three orders of magnitude smaller than the QCD counterpart in the wide range of temperatures explored here, to wit. The ratio $R$ of this hadronic scattering term, Eq. (9), and its QCD counterpart, Eq.(5), is $R\simeq 10^{-3}$ at $T= 100 \;\mbox{MeV}$, and $R\simeq 10^{-2}$ at $T= 160 \;\mbox{MeV}$.\\

The correlation function $\Pi(q^2,T)$, Eq.(1), satisfies a once subtracted dispersion relation. To eliminate the subtraction one can use Hilbert moments, i.e.

\begin{equation}
\varphi_N(Q^2,T) \equiv \frac{(-)^{N}}{(N)!}\, \Bigl(\frac{d}{dQ^2}\Bigr)^{N} \Pi(Q^2,T)\, = \frac{1}{\pi}
\int_{0}^{\infty} \; \frac{ds}{(s+Q^2)^{N+1}}\,  Im \,\Pi(s,T)\; , 
\end{equation}

where $N = 1,2,...$, and $Q^2 \geq 0$ is an external four-momentum squared, to be considered as a free parameter as discussed below. 
Using Cauchy's theorem in the complex s-plane, which is equivalent to invoking quark-hadron duality, the Hilbert moments become Finite Energy QCD sum rules (FESR), i.e.
\begin{equation}
\varphi_N(Q^2, T)|_{RES} =  \varphi_N(Q^2,T)|_{QCD} \;,
\end{equation}
where
\begin{equation}
\varphi_N(Q^2,T)|_{RES} \equiv \frac{1}{\pi}
\int_{0}^{s_0(T)}\frac{ds}{(s+Q^2)^{N+1}}\, Im \,\Pi(s,T)|_{RES} \;,
\end{equation}

\begin{eqnarray}
 \varphi_N(Q^2,T)|_{QCD} &\equiv& \frac{1}{\pi}
\int_{4 m_Q^2}^{s_0(T)}\frac{ds}{(s+Q^2)^{N+1}}\, Im \,\Pi_{a}(s,T) \nonumber \\ [.3cm]
&+& \frac{1}{\pi} \int_{0}^{\infty}\frac{ds}{(s+Q^2)^{N+1}}\, Im \,\Pi_{s}(s,T) 
+ \varphi_N(Q^2,T)|_{NP}  \;,
\end{eqnarray}
and  $Im \,\Pi(s,T)|_{RES}$ is given by the first term in Eq.(6) modified in finite-width according to Eq.(8), and the PQCD spectral functions are given by  Eqs.(4) and (5). The non-perturbative term corresponding to the dimension d=4 in the OPE is given by \cite{QCDSR}
\begin{eqnarray}
\varphi_N(Q^2,T)|_{NP} &=& - \frac{3}{4\pi^2}\frac{1}{(4m_Q^2)^N}\frac{1}{(1+\xi)^{N+2}}\;F\left(N+2,\,-\frac{1}{2}\,,N+\frac{7}{2},\,\rho\right) \nonumber \\ [.3cm]
&\times& \frac{2^N \,N\,(N+1)^2\,(N+2)\,(N+3)\,(N-1)!}{(2N+5)\,(2N+3)!!}\;\Phi
\end{eqnarray}
where $F(a,b,c,z)$ is the hypergeometric function, , $\xi\equiv\frac{Q^2}{4m_Q^2}$, $\rho\equiv\frac{\xi}{1+\xi}$, and
\begin{equation}
\Phi\equiv\frac{4\pi^2}{9}\frac{1}{(4m_Q^2)^2}\left<\left<\frac{\alpha_s}{\pi}G^2\right>\right>\;,
\end{equation}
\begin{figure}[ht]
\begin{center}
\includegraphics[width=\columnwidth]{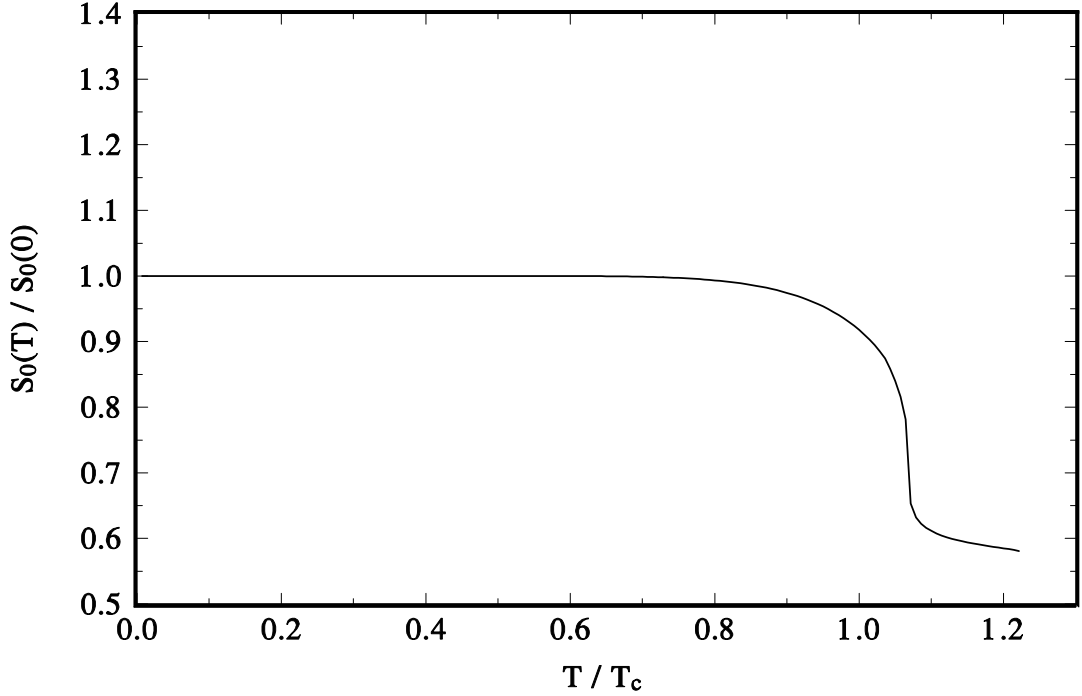}
\caption{The  ratio $s_0(T)/s_0(0)$  as a function of $T/T_c$.}
\end{center}
\end{figure}

\begin{figure}[ht]
\begin{center}
\includegraphics[width=\columnwidth]{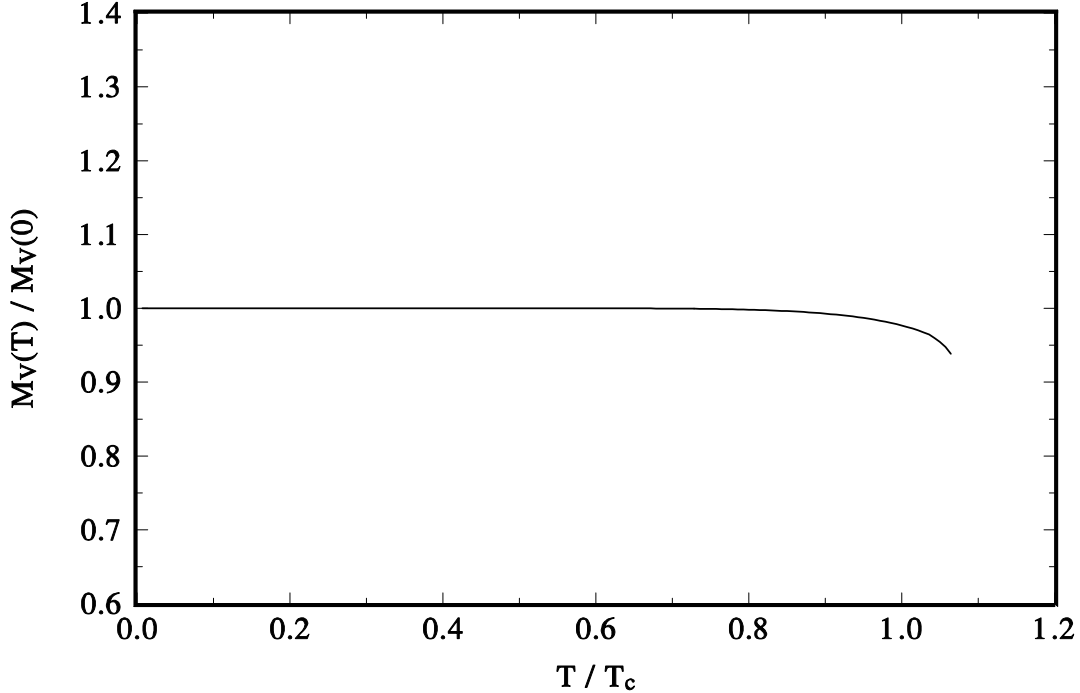}
\caption{The  ratio $M_V(T)/M_V(0)$ as a function of $T/T_c$. This ratio is basically the same in zero-width as in finite width.}
\end{center}
\end{figure}
The gluon condensate  $\left<\left<\frac{\alpha_s}{\pi}G^2\right>\right>$ at low temperatures has been calculated in chiral perturbation theory \cite{BERN} with the following result
\begin{equation}
\left<\left<\frac{\alpha_s}{12\, \pi} G^2\right>\right>= \left<\frac{\alpha_s}{12\, \pi} G^2\right> -\, \frac{\alpha_s}{\pi} \,\frac{\pi^4}{405}\, \frac{N_F^2 (N_F^2-1)}{33- 2 N_F} (\ln \frac{\Lambda_p}{T} - 1) \frac{T^8}{f_\pi^4}\;,
\end{equation}
where $N_F$ is the number of quark flavours, and $\Lambda_p \approx 200 - 400 \;\mbox{MeV}$. To a good approximation this can be written as
\begin{equation}
\left<\left<\frac{\alpha_s}{12\, \pi} G^2\right>\right>= \left<\frac{\alpha_s}{12\, \pi} G^2\right> \Big[ 1 - \Big (\frac{T}{T_c}\Big)^8\Big] \;.
\end{equation}
Due to this thermal behaviour the gluon condensate, in this framework, remains essentially constant up to temperatures close to $T_c \simeq 100 \; \mbox{MeV}$, after which it decreases very sharply. In order to go beyond the low temperature regime of chiral perturbation theory, lattice QCD  provides the right tool. A good approximation to results in this framework \cite{latticeG} is given by the expression
\begin{equation}
\left<\left<\frac{\alpha_s}{\pi}G^2\right>\right> = \left<\frac{\alpha_s}{\pi}G^2\right> \left[\theta(T^*-T) + \frac{1-\frac{T}{T_C^*}}{1-\frac{T^*}{T_C^*}}\theta(T-T^*) \right]
\end{equation}
where $T^*\approx 150$ MeV is the breakpoint temperature where the condensate begins  to decrease appreciably, and $T_C^*\approx 250$ MeV is the temperature at which $\left<\left<\frac{\alpha_s}{\pi}G^2\right>\right>_{T_C}=0$.\\ 
Returning to the $Q^2$ dependence of the Hilbert moments, Eq.(10), it has been customary in many analyses of charmonium at T=0  to take $Q^2=0$, but the case $Q^2 > 0$ has also been advocated \cite{RRY}. At finite temperature, though, due to the singular behaviour of the space-like PQCD term, Eq.(5), one is compelled to take $Q^2 > 0$. Since all hadronic parameters will be normalized to their values at T=0, and we are only interested here in their temperature behaviour, we shall fix $Q^2$ as well as $s_0(0)$ from the experimental values of the mass, the coupling, and the width at T=0.
At finite temperature there are additional contributions to the OPE in the form of non-diagonal (Lorentz non-invariant) condensates.  In the case of non-gluonic operators they are highly suppressed \cite{HL}, \cite{ELE} so that they can be safely ignored. A gluonic twist-two term in the OPE has been considered in \cite{GLUON2}, and computed on the lattice in \cite{MORITA}. Using this information we find that the non-perturbative QCD moments, $\varphi_N(Q^2,T)|_{NP}$, change as follows
\begin{equation}
\varphi_N(Q^2,T)|_{NP} \rightarrow \varphi_N(Q^2,T)|_{NP} \left[1\;+\;\delta_N(Q^2,T)\right]\;,
\end{equation}
where
\begin{equation}
\delta_N(Q^2,T) = \left[ 1 \,+ \,\frac{4}{3}\; \frac{1}{(N+2)(N+3)}\; \frac{F(N+2,-3/2,N+7/2;\rho}{F(N+2,-1/2,N+7/2;\rho}\right] \;\left(3 \;\frac{G_2(T)}{G_0(T)}\right)\;,
\end{equation}
with $G_0(T) \equiv <<\frac{\alpha_s}{\pi} \,G^2>>$, $G_0 (0)= (0.05 \pm 0.02)\;\mbox{GeV}^4$ \cite{G2},  $G_2(T) \simeq - 10^{-3} \;\mbox{GeV}^4$ \cite{MORITA}, in the range of temperatures considered here, and the parameters $\xi$ and $\rho = (1/2 - 1)$ have been defined after Eq.(14).
For small values of $N$ ($N = 1-3$) the second term in brackets above is at the level of a couple of percent, while for larger values of $N$ it becomes negligible. The correction term is small and essentially independent of $Q^2$, i.e. $\delta_N(Q^2,T) \simeq (2 - 6)\%$  in the temperature range considered here, and will play no appreciable role in the results, as will be discussed later.

\section{Results}
\noindent
We begin by determining $s_0$ and $Q^2$ at T=0 from the moments, Eq.(11), and using as input the experimental values \cite{PDG} $M_V = 3.097 \;\mbox{GeV}$, $f_V = 196 \;\mbox{MeV}$, and $\Gamma_V = 93.2 \; \mbox{keV}$, as well as $m_Q = 1.3 \;\mbox{GeV}$, and  \cite{G2} $ \langle0| \frac{\alpha_s}{12 \pi} G^2 |0\rangle \simeq 5 \times 10^{-3} \mbox{GeV}^4$. In the zero-width approximation one finds from Eq.(12) that
\begin{equation}
\frac{\varphi_1(Q^2)|_{RES}}{\varphi_2(Q^2)|_{RES}} = \frac{\varphi_2(Q^2)|_{RES}}{\varphi_3(Q^2)|_{RES}}\;.
\end{equation} 

\begin{figure}[ht]
\begin{center}
\includegraphics[width=\columnwidth]{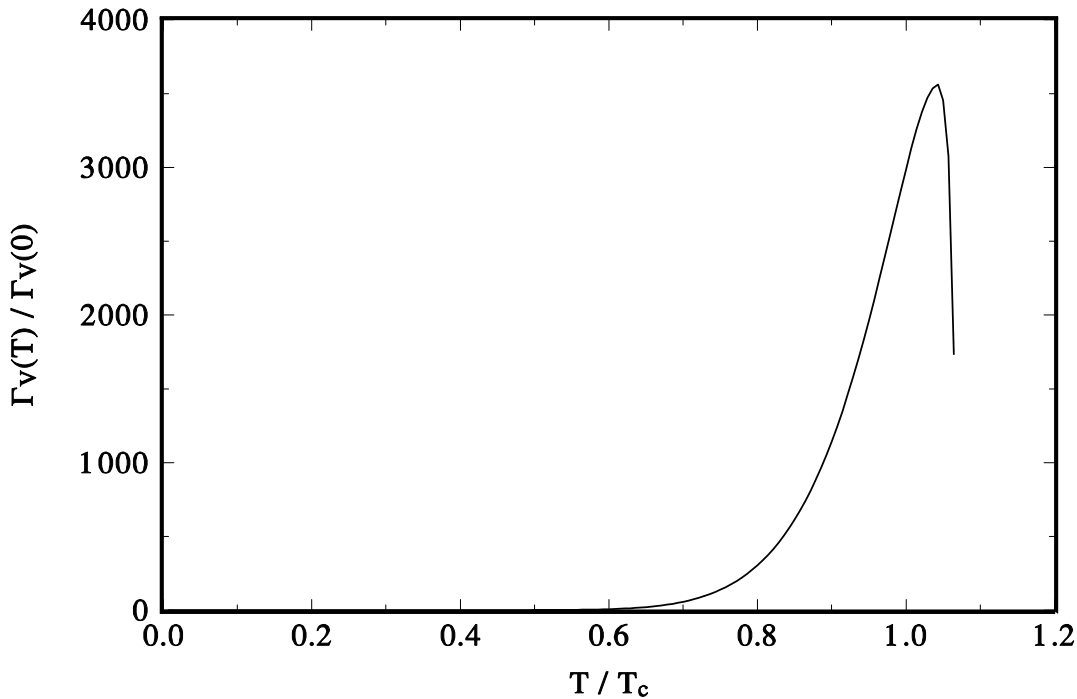}
\caption{The  ratio $\Gamma_V(T)/\Gamma_V(0)$  as a function of $T/T_c$.}
\end{center} 
\end{figure}

Given the extremely small total width of the $J/\psi$ it turns out that the above relation also holds with extreme accuracy in finite width. Using Eq.(11) this leads to
\begin{equation}
\frac{\varphi_1(Q^2)|_{QCD}}{\varphi_2(Q^2)|_{QCD}} = \frac{\varphi_2(Q^2)|_{QCD}}{\varphi_3(Q^2)|_{QCD}}\;,
\end{equation} 
which depends only on the two unknowns $s_0$ and $Q^2$, and provides the first equation to determine this pair of parameters. The second equation can be e.g. Eq.(11) with $N=1$. In this way we find that $s_0 = 11.64 \;\mbox{GeV}^2$, and $Q^2 = 10 \;\mbox{GeV}^2$ reproduce the experimental values of the mass, coupling, and width of $J/\psi$ within less than 1\%. We have checked that equally reasonable agreement for the hadronic parameters at $T=0$ can be achieved using larger values of $Q^2$, up to $Q^2 \simeq 20 \; \mbox{GeV}^2$, as well as larger values of $N$, up to say $N \simeq \, 10$. These different choices do not produce any qualitative change at finite $T$, other than some $10 \,\%$ shift of the critical temperature, upwards for larger $Q^2$, and downwards for larger $N$. 
This whole set of hadronic parameters will then be used to normalize the corresponding parameters at finite temperature. The latter are obtained as follows. \\
\begin{figure}[ht]
\begin{center}
\includegraphics[width=\columnwidth]{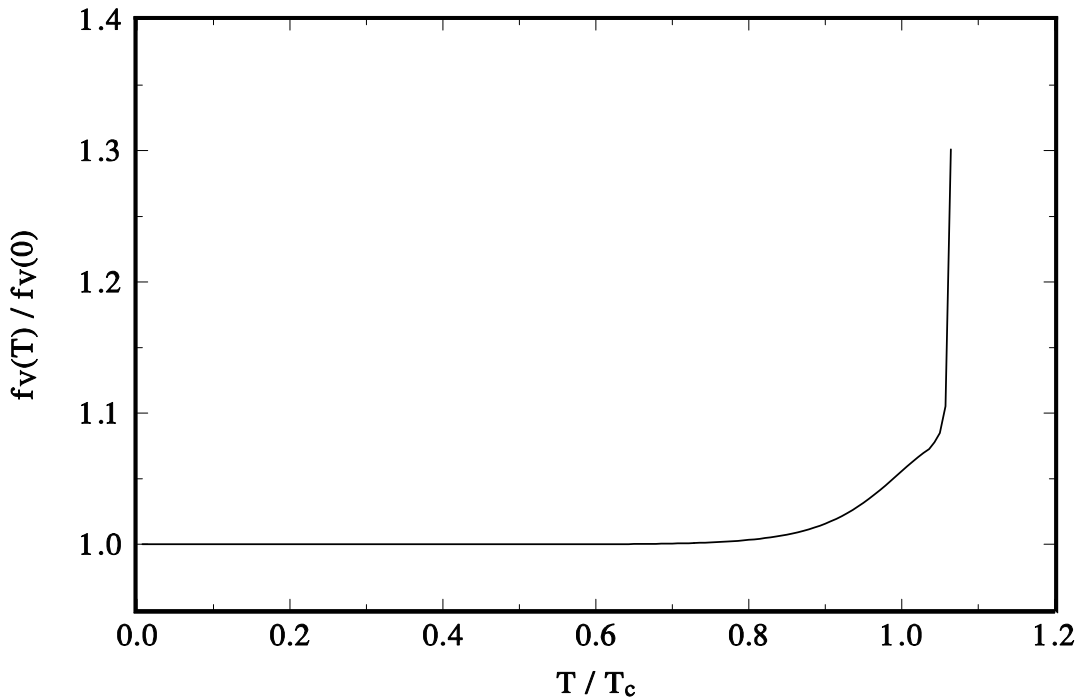}
\caption{The  ratio $f_V(T)/f_V(0)$ as a function of $T/T_c$.}
\end{center}
\end{figure}
The continuum threshold $s_0(T)$ is obtained again using Eq.(22), as Eq.(21) continues to hold to great accuracy in finite width, even if the width were to increase with temperature by three orders of magnitude, say from $\Gamma_V(0) \simeq 0.1 \; \mbox{MeV}$ to $\Gamma_V(T) \simeq 300 \; \mbox{MeV}$. The result for $s_0(T)$ normalized to $s_0(0)$ is shown in Fig.1 as a function of $T/T_c$, where $T_c \simeq 140 \;\mbox{MeV}$ is the temperature where $s_0(T)$ begins to deviate appreciably from $s_0(0)$. Above this temperature $s_0(T)$ decreases sharply until it approaches the threshold $s_{th} = 4 \, m_Q^2$ at a temperature $T  \simeq\; 1.22 \, T_c$, with this threshold being the minimum possible value for $s_0(T)$. It must be stressed that the temperature dependence of the continuum threshold is rather important in order to avoid inconsistencies, and therefore it should not be ignored. In the case of light-light and heavy-light quark systems $s_0(T)$ is related to the thermal light quark condensate. This is not the case for heavy-heavy quark hadrons, so that $s_0(T)$ must be determined from the Hilbert moments together with the other hadronic parameters.
Next, the mass, in the zero-width approximation, follows from the ratio
\begin{equation}
\frac{\varphi_1(Q^2,T)|_{RES}}{\varphi_2(Q^2,T)|_{RES}} = \frac{\varphi_1(Q^2,T)|_{QCD}}{\varphi_2(Q^2,T)|_{QCD}}\;.
\end{equation} 
The result for $M_V(T)/M_V(0)$ as a function of $T/T_c$ is shown in Fig.2. The mass remains basically constant until $T \simeq \, T_c$ where it decreases slightly by a few percent. Beyond $T  \simeq \, 1.1 \, T_c$ the whole program of determining the temperature behaviour of the hadronic parameters breaks down. The simple reason for this is that at this temperature $s_0(T)$ has left the resonance behind, i.e. $s_{th} < M_V^2$. Hence, there is no longer a support for the Hilbert moments in the hadronic sector.
We have verified that this result for $M_V(T)/M_V(0)$ remains essentially the same in finite width. Hence, in determining the coupling and the width we shall keep the mass independent of the temperature. Next, the width can be determined from the ratio  
\begin{equation}
\frac{\varphi_1(Q^2,T)|_{RES}}{\varphi_3(Q^2,T)|_{RES}} = \frac{\varphi_1(Q^2,T)|_{QCD}}{\varphi_3(Q^2,T)|_{QCD}}\;.
\end{equation} 
The result for $\Gamma_V(T)/\Gamma_V(0)$  as a function of $T/T_c$ is shown in Fig. 3; the peak is reached at $T \simeq \, T_c$. Finally, the leptonic decay constant can be determined from any single moment, e.g. from
\begin{equation}
\varphi_1(Q^2,T)|_{RES} = \varphi_1(Q^2,T)|_{QCD}.
\end{equation}
The ratio $f_V(T)/f_V(0)$ as a function of $T/T_c$ is shown in Fig. 4. The rise of the coupling beyond $T_c$ is the result of a detailed balance between different contributions. With increasing $T$ the continuum threshold, $s_0(T)$, decreases and so does the PQCD annihilation moment, while the QCD scattering moment increases approaching the annihilation moment at $T\simeq 160 \;\mbox{MeV}$; all the while, the non perturbative moment remains approximately constant and negative, and numerically comparable to the other two terms at that temperature, thus mostly canceling the annihilation  moment there. The hadronic moment dividing the QCD moment is mostly a decreasing function of the temperature due to the decrease of $s_0(T)$ and to the increase of the width, leading finally to an increase in the coupling. This behaviour of the coupling and width of the $J/\psi$ 
remains qualitatively stable against changes in $N$ and $Q^2$ in the moments, for $Q^2 \simeq 10-20 \;\mbox{GeV}^2$, and $N \simeq 1 - 10$. Numerically, higher values of $Q^2$ tend to increase slightly the critical temperature, and higher values of $N$ tend to reduce it. We have explicitly verified that the hadronic scattering moment can be safely ignored.
Both the width and the coupling can only be determined up to $T_f \simeq\; 1.1 \, T_c$ beyond which $s_0(T) < M_V^2(T)$. The temperature behaviour of the width and the coupling thus obtained strongly suggests the  survival of the $J/\psi$ above the critical temperature for deconfinement.\\
\section{Discussion and conclusions}
\noindent
In this paper we have determined the temperature behaviour of the hadronic parameters of the $J/\psi$ resonance using Hilbert moment QCD sum rules. The hadronic parameters are not only the mass, coupling (leptonic decay constant), and total width, but also the continuum threshold $s_0$ which signals the onset of  PQCD. This threshold at finite temperature was first introduced in \cite{BOCH}, where it was interpreted as a phenomenological parameter signaling deconfinement. Later, it was related  to  chiral symmetry restoration \cite{CAD3}, and a scaling law was found associating $s_0(T)$ to the thermal (light) quark condensate \cite{radii}. Hence, this parameter constitutes an essential part of any QCD sum rule analysis at finite $T$, and ignoring the temperature dependence of $s_0$ would lead to inconsistencies. The results of our analysis at low temperatures show that the $J/\psi$ behaves as other light-light and heavy-light quark resonances, i.e. the continuum threshold and the coupling decrease and the width increases with increasing $T$. However, as $T$ approaches $T_c$  this behaviour changes so that the coupling increases and the width decreases. The PQCD spectral function in the space-like region (scattering term), as well as the gluon condensate, are responsible for this new scenario which strongly suggests the survival of the $J/\psi$ beyond deconfinement. This is a feature unique to heavy-heavy quark systems, as this space-like contribution  is negligible in light-light and even heavy-light quark channels \cite{HL}. While this PQCD term increases monotonically in importance with increasing $T$, the QCD sum rule framework necessarily breaks down at the temperature where $s_0(T)$ reaches the minimum  value $s_0|_{min} = M^2_{J/\psi}$. In fact, since the hadronic mass is essentially independent of $T$, once  $s_0(T) < M_{J/\psi}^2$ there is no further support for the Hilbert integrals. In spite of this, one could continue to explore the hadronic parameter space by reducing arbitrarily the value of $M_{J/\psi}$ at $T=0$. In this case, one would find that the behaviour of the coupling and width persists at higher temperatures.
Our results are not necessarily in conflict with a recent QCD sum rule analysis   \cite{MORITA}, where charmonium was found to survive beyond $T_c$. The reason is that in  \cite{MORITA} the scattering term was not taken into account, and the temperature dependence of the continuum threshold was not considered. Both these features are essential for a consistent QCD sum rule analysis. Studies of charmonium at finite $T$ in non-relativistic frameworks \cite{MOCSY}-\cite{BARI} also lead to charmonium melting at some critical temperature. However, in the non-relativistic regime the complex energy plane has only a right-hand cut extending from zero to infinity, which corresponds to the time-like case in the relativistic domain. The space-like cut at finite $T$ centered at the origin in the complex energy plane, and covering the range $ -|\bf{q}| \leq \omega \leq + |\bf{q}|$, is a genuine relativistic effect. The results and conclusions of the present analysis, though,  are in agreement with those obtained from QCD numerical simulations on the lattice \cite{lattice}. \\
\section{Appendix}
\noindent
In this appendix we outline the calculation of the imaginary part of the correlator, Eq.(1), in PQCD and in the space-like region, as there are some incorrect results in the literature. To one-loop order in PQCD this is given by \cite{BOCH}

\begin{equation}
\frac{1}{\pi}\,\mbox{Im}\; \Pi_s(q^2,T) = \frac{3}{8 \pi^2} \int_v^{\infty} dx \, (1 - x^2)   \;\Delta n_F(x) 
\end{equation}

where
\begin{equation}
 \Delta n_F(x) \equiv n_F\left(\frac{|\mathbf{q}| x + \omega}{2T}\right) - n_F\left(\frac{|\mathbf{q}| x - \omega|}{2T}\right) \;.
\end{equation}

Performing the change of variable $ y = \frac{|\mathbf{q}| x}{2T}$,
the Fermi factor $\Delta n_F$ in Eq.(26) becomes the total derivative

\begin{equation}
\Delta n_F = \frac{\omega}{T} \;\frac{d}{d y} \; n_F(y) \;.
\end{equation}

After integration by parts, Eq.(26) becomes

\begin{equation}
Im \Pi_s(q^2,T) =  \frac{3}{4 \pi} \, \frac{\omega}{|\bf{q}|} \, 
\left[- n_F \left(\frac{|\mathbf{q}| v}{2T}\right) (1 - v^2) + \frac{8 T^2}{|\mathbf{q}|^2} \int_{\frac{|\mathbf{q}| v}{2T}}^{\infty} \;y \; n_F(y) \;dy \right]\;.
\end{equation}

Finally, taking the limit $\omega \rightarrow 0$, followed by $|\bf{q}| \rightarrow 0$, and using the expression

\begin{equation}
\lim_{|\bf{q}| \rightarrow 0}\; \lim_{\omega \rightarrow 0} \left(\frac{\omega}{|\bf{q}|^3}\right) = \frac{2}{3} \; \delta\left( \omega^2 \right)\;,
\end{equation}

gives the final result for the PQCD spectral function in the space-like region, i.e.

\begin{equation}
\frac{1}{\pi}\, Im \,\Pi_s(\omega,T) =  \frac{2}{ \pi^2}\; m_Q^2\; \delta(\omega^2)\left[ n_F \left(\frac{m_Q}{T}\right)\;+ \frac{2\,T^2}{m_Q^2} \int_{m_Q/T }^{\infty} y \,n_F (y) \, dy \;\right]
\;.
\end{equation}


\begin{thebibliography}{99}
\bibitem{QCDSR} For a recent review see e.g. P. Colangelo and A. Khodjamirian, in: "At the Frontier of Particle Physics/ Handbook of QCD", M. Shifman, ed. (World Scientific, Singapore 2001), Vol. 3, 1495-1576.

\bibitem{BOCH} A.I. Bochkarev and M.E. Shaposnikov, Nucl. Phys. B 286 (1986) 220.

\bibitem{CAD1} C.A. Dominguez and M. Loewe, Physical Review  D 52 (1995) 3143.

\bibitem{VARIOUS} R.J. Furnstahl, T. Hatsuda and S.H. Lee,  Phys. Rev. D 42 (1990) 1744;
C. Adami, T. Hatsuda and I. Zahed, Phys. Rev. D 43(1991) 921; C. Adami and I. Zahed, Phys. Rev. D 45 (1992) 4312; T. Hatsuda, Y. Koike and S.-H. Lee, Phys. Rev. D 47 (1993) 1225; {\it ibid.} Nucl. Phys. B 394 (1993) 221; Y. Koike, Phys. Rev. D 48 (1993) 2313. 

\bibitem{CAD2} C.A. Dominguez and M. Loewe, Z. Phys. C (Particles \& Fields) 51 (1991) 69; {\it ibid.} 58 (1993) 273; Phys. Lett. B 481 (2000) 295.

\bibitem{HL} C.A. Dominguez, M. Loewe and J.C. Rojas, J. High Energy Phys. 0708 (2007) 040; E.V. Veliev and G. Kaya, arXiV:0902.3443.

\bibitem{CAD0} C.A. Dominguez and M. Loewe, Z. Phys. C (Particles \& Fields) 49 (1991) 423.

\bibitem{radii} C.A. Dominguez, M. Loewe and J.S. Rozowsky, Phys. Lett. B 335 (1994) 506; C.A. Dominguez, M. S. Fetea and M. Loewe, Phys. Lett.  B 387 (1996) 151; {\it ibid} B 406 (1997) 149; C.A. Dominguez, M. Loewe and  C. van Gend, Phys. Lett.  B 429 (1998) 64; {\it ibid} B 460 (1999) 442.  

\bibitem{CAD3} C.A. Dominguez and M. Loewe, Phys. Lett. B 233 (1989) 201.
The (near) equality of the critical temperatures for chiral-symmetry restoration and
deconfinement was shown analytically in A. Barducci, R. Casalbuoni, S. De Curtis, R. Gatto
and G. Pettini, Phys. Lett. B 244 (1990) 311. These authors used a result for the thermal quark condensate valid for $ 0 \leq T \leq T_c$, while the first reference only made use of the low-T expansion of chiral perturbation theory, obtaining somewhat different critical temperatures.

\bibitem{lattice} For recent results see e.g. H. Ohno {\it et al.}, arXiv:08103066 and references therein.

\bibitem{mQ} T. Altherr and D. Seibert, Phys. Rev. C 49 (1994) 1684.

\bibitem{BERN} P. Gerber and H. Leutwyler, Nucl. Phys. B 321 (1989) 387.

\bibitem{latticeG} G. Boyd and D. E. Miller, arXiv:hep-ph/9608482 (unpublished); D.E. Miller, arXiv:hep-ph/0008031 (unpublished)

\bibitem{RRY} L.J. Reinders, H. Rubinstein and S. Yazaki, Phys. Rep. 127 C (1985) 1.

\bibitem{ELE} V.L. Eletsky, Phys. Lett. B 352 (1995) 440.

\bibitem{GLUON2} F. Klingl, S. Kim, S.H. Lee, P. Morath, and W. Weise, Phys. Rev. Lett. 82, 3396 (1999).

\bibitem{MORITA} K. Morita and S.H. Lee, Phys. Rev. Lett. 100 (2008) 022301; arXiv:0711.3998.

\bibitem{G2}   R.A. Bertlmann, {\it et al.}, Z. Phys. C (Particles \& Fields)  39 (1988) 231; C.A. Dominguez and J. Sola, Z. Phys. C (Particles \& Fields) 40 (1988) 63; C.A. Dominguez and K. Schilcher, J. High Energy Phys. 01 (2007) 093.   
 
\bibitem{PDG}  Particle Data Group, C. Amsler {\it et al.}, Phys. Lett. B 667, 1 (2008).

\bibitem{MOCSY} For a recent review see e.g. A. Mocsy, arXiv:0908.0746, and references therein.

\bibitem{BARI} For a recent analysis see e.g. F. Giannuzzi and M. Mannarelli, arXiv:0907.1041, and references therein.

\end{thebibliography}
\end{document}